\documentstyle[prd,epsf,aps]{revtex}

\newcommand{\thh}{(\frac{\pi}{2}-\theta)}

\begin{document}

\title{Static Axially Symmetric Einstein-Yang-Mills-dilaton solutions:
Addendum Asymptotic solutions}
\vspace{1.5truecm}
\author{
{\bf Burkhard Kleihaus}}
\address{NUIM, Department of Mathematical Physics, Maynooth, Co. Kildare,
Ireland\\
}
\author{and}
\author{
{\bf Jutta Kunz}}
\address{Fachbereich Physik, Universit\"at Oldenburg, D-26111 Oldenburg,
Germany
}
\date{\today}
\maketitle

\begin{abstract}
We discuss the asymptotic form of
the static axially symmetric,
globally regular and black hole solutions,
obtained recently in Einstein-Yang-Mills and 
Einstein-Yang-Mills-dilaton theory.
\end{abstract}
\vfill
\noindent {Preprint gr-qc/9909160} \hfill\break
\vfill\eject
%\vspace{1.5truecm}

Recently, we have constructed static axially symmetric
regular and black hole solutions
in SU(2) Einstein-Yang-Mills (EYM) 
and Einstein-Yang-Mills-dilaton (EYMD) theory
\cite{kk1,kk2,kk3,kk4,regu,gv}.
Representing generalizations
of the spherically symmetric regular and black hole solutions
\cite{bm,su2,eymd,gv},
these solutions are characterized by two integers,
their winding number $n$ and
the node number $k$ of their gauge field functions.
The spherically symmetric solutions have winding number $n=1$.

These non-abelian EYM and EYMD solutions are asymptotically flat.
They have non-trivial magnetic gauge field configurations,
but carry no global charge.
To every globally regular solution
there exists a corresponding family of black hole solutions
with regular event horizon $x_{\rm H}>0$.
These non-abelian black hole solutions demonstrate
that neither the ``no-hair'' theorem nor Israel's theorem
hold in EYM and EYMD theory.

Here we give a detailed account of the 
asymptotic form of these static axially symmetric solutions.
In particular we find, that the expansion of the gauge field functions 
in powers of $1/x$ must be supplemented with
non-analytic terms for $n=2$ and 4.

Let us briefly recall the SU(2) Einstein-Yang-Mills-dilaton action
\begin{equation}
S=\int \left ( \frac{R}{16\pi G} + L_M \right ) \sqrt{-g} d^4x
\ \label{action} \end{equation}
with matter Lagrangian
\begin{equation}
L_M=-\frac{1}{2}\partial_\mu \Phi \partial^\mu \Phi
 -e^{2 \kappa \Phi }\frac{1}{2} {\rm Tr} (F_{\mu\nu} F^{\mu\nu})
\ , \label{lagm} \end{equation}
field strength tensor
$
F_{\mu \nu} = 
\partial_\mu A_\nu -\partial_\nu A_\mu + i e \left[A_\mu , A_\nu \right] 
$,
gauge field
$ A_{\mu} = \frac{1}{2} \tau^a A_\mu^a $,
dilaton field $\Phi$,
and Yang-Mills and dilaton coupling constants
$e$ and $\kappa$, respectively.

In terms of the polar coordinates $r$, $\theta$ and $\phi$
the isotropic metric reads \cite{kk3,kk4}
\begin{equation}
ds^2=
  - f dt^2 +  \frac{m}{f} d r^2 + \frac{m r^2}{f} d \theta^2 
           +  \frac{l r^2 \sin^2 \theta}{f} d\phi^2
\ , \label{metric2} \end{equation}
where $f$, $m$ and $l$ are only functions of $r$ and $\theta$,
and regularity on the $z$-axis requires
$m|_{\theta=0}=l|_{\theta=0}$ \cite{kk3,kk4}.

We parameterize the static axially symmetric gauge field as
\cite{kk1,kk2,kk3,kk4}
\begin{equation}
A_\mu dx^\mu =
\frac{1}{2er} \left[ \tau^n_\phi 
 \left( H_1 dr + \left(1-H_2\right) r d\theta \right)
 -n \left( \tau^n_r H_3 + \tau^n_\theta \left(1-H_4\right) \right)
  r \sin \theta d\phi \right]
\ , \label{gf1} \end{equation}
where $n$ denotes the winding number,
the $su(2)$ matrices $\tau^n_\varphi,\tau^n_r,\tau^n_\theta$ are 
defined in terms of Pauli matrices $\tau_1,\tau_2, \tau_3$ by
\begin{equation}
\tau^n_\varphi = -\sin(n\varphi) \tau_1+\cos(n\varphi) \tau_2 \ ,
\tau^n_r = \sin\theta \tau^n_\rho + \cos\theta \tau_3 \ ,
\tau^n_\theta = \cos\theta \tau^n_\rho - \sin\theta \tau_3 \ ,
\end{equation}
with $\tau^n_\rho= \cos(n\varphi) \tau_1+\sin(n\varphi) \tau_2$.
The $H_i$ are only functions of $r$ and $\theta$,
and regularity on the $z$-axis requires 
$ H_2|_{\theta=0}=H_4|_{\theta=0}$.

Under abelian gauge transformations \cite{kk1,kk2,kk3,kk4,regu}
\begin{equation}
U = \exp{\left\{i \Gamma \tau^n_\varphi/2\right\}} \ ,
\label{gam}
\end{equation}
where $\Gamma$ is a function of $r$ and $\theta$,
the gauge potential (\ref{gf1}) is form invariant and
the functions $H_i$ transform like
\begin{eqnarray}
{H_1} &\longrightarrow &\hat{H}_1 
= H_1 - r \partial_r \Gamma  \ ,
              \label{ET_H1}\\
{H_2} &\longrightarrow &\hat{H}_2 
= H_2 +  \partial_\theta \Gamma  \ ,
              \label{ET_H2}\\
{H_3} &\longrightarrow & \hat{H}_3 
= \cos \Gamma (H_3+\cot \theta) -\sin \Gamma H_4 - \cot \theta 
             \ ,   \label{ET_H3}\\
{H_4} &\longrightarrow & \hat{H}_4 
= \sin \Gamma (H_3+\cot \theta) +\cos \Gamma H_4 
             \ .  \label{ET_H4}\\
\nonumber
\end{eqnarray}
We fix the gauge by choosing \cite{kk1,kk2,kk3,kk4} 
\begin{equation}
 r \partial_r H_1 - \partial_\theta H_2 = 0 
\ . \label{gc1} \end{equation}

For convenience we introduce dimensionless quantities
\cite{kk1,kk2,kk3,kk4}
\begin{equation}
x=\frac{e}{\sqrt{4\pi G}} r
\ , \ \ \
\varphi = \sqrt{4\pi G} \Phi
\ , \ \ \
\gamma =\frac{1}{\sqrt{4\pi G}} \kappa
\ . \label{dimless} \end{equation}

To obtain globally regular solutions or 
black hole solutions with a regular horizon
with the proper symmetries,
we must impose appropriate boundary conditions \cite{kk1,kk2,kk3,kk4}.
For asymptotically flat,
magnetically neutral solutions the boundary conditions 
at infinity are
\begin{equation}
f|_{x=\infty}= m|_{x=\infty}= l|_{x=\infty}=1
\ , \label{bc1a} \end{equation}
\begin{equation}
H_2|_{x=\infty}=H_4|_{x=\infty}=\pm 1 \ , \ \ \ 
H_1|_{x=\infty}=H_3|_{x=\infty}=0 \ , \ \ \
\label{bc1c} \end{equation}
and we fix the scale invariance of the field equations by the condition
$\varphi|_{x=\infty}=0$.

We now consider the asymptotic form of the functions at infinity.
In Appendix C of \cite{kk3}
we presented the expansion of the functions 
at infinity in powers of $1/x$, not considering
possible non-analytic terms.
This expansion reads \cite{kk3}
\begin{eqnarray}
H_1 & = & 
\frac{1}{x^2} \bar{H}_{12} \sin \theta \cos \theta
 + O\left(\frac{1}{x^3}\right)\ ,
\nonumber \\ 
H_2 & = & 
\pm 1 + \frac{1}{2 x^2} \bar{H}_{12} 
\left( \cos^2 \theta - \sin^2 \theta \right)
 + O\left(\frac{1}{x^3}\right)\ ,
\nonumber \\ 
H_3 & = & 
\frac{1}{x} \sin \theta \cos \theta \bar{H}_{31}
-\frac{1}{4 x^2} \sin \theta \cos \theta \left( \pm 2 \bar{H}_{12} 
+ \bar{H}_{31} \left( \bar{f}_1 + 2 \gamma \bar{\varphi}_1 \right) \right)
 + O\left(\frac{1}{x^3}\right)\ ,
\nonumber \\ 
H_4 & = & 
\pm \left( 1+ \frac{1}{x} \bar{H}_{31} \sin^2 \theta 
          \pm \frac{1}{2 x^2} \bar{H}_{12}
         -\frac{1}{4 x^2} \sin^2 \theta 
         \left( 
         \pm 2 \bar{H}_{12} 
         + \bar{H}_{31} \left( \bar{f}_1 + 2 \gamma \bar{\varphi}_1 \right)
         \right)
         \right) 
 + O\left(\frac{1}{x^3}\right)\ ,
\nonumber \\ 
\varphi & = & \frac{1}{x} \bar{\varphi}_1 
 + O\left(\frac{1}{x^3}\right)\ ,
\nonumber \\ 
f & = & 
1+ \frac{1}{x} \bar{f}_1 + \frac{1}{2 x^2} \bar{f}_1^2
 + O\left(\frac{1}{x^3}\right)\ ,
\label{mfold} \\ 
m & = &
1+ \frac{1}{x^2} \bar{l}_2 + \frac{1}{x^2} \sin^2 \theta \bar{m}_2
 + O\left(\frac{1}{x^3}\right)\ ,
\label{mmold} \\
l & = &
1 + \frac{1}{x^2} \bar{l}_2
 + O\left(\frac{1}{x^3}\right)\ ,
\label{mlold} 
\end{eqnarray}
with constants $\bar{H}_{12}$, $\bar{H}_{31}$, $\bar{f}_1$, $\bar{\varphi}_1$,
$\bar{m}_2$ and $\bar{l}_2$.

The above expansion in powers of $1/x$ is, however, not necessarily
complete, since non-analytic terms may be present.
Inspection of the numerical solutions \cite{kk1,kk2,kk3,kk4} reveals,
that $\ln{x}$ terms must be included in the expansion
of the gauge field functions
for $n=2$ and 4, to obtain the proper asymptotic form of the solutions.
In particular, 
for an even number of nodes and for winding number $n=2$
we obtain
\begin{eqnarray}
H_1 & = & 
\frac{ D_2\ln{x}\sin2 \theta}{2 x^2}
+\frac{1}{x^2}\left\{
\frac{D_2}{4} \left[2 \thh \cos 2\theta + \cos\theta (\sin\theta-\pi)\right]
-D_0 \sin 2\theta\right\} \nonumber \\
& &
+ {\rm higher}\ {\rm order}\ {\rm terms} \ ,
\label{eqn2h1}\\
& & \nonumber \\
H_2 & = &
1+\frac{ D_2\ln{x}\cos 2\theta}{2 x^2}
+\frac{1}{x^2}\left\{
\frac{D_2}{8} \left[\cos 2\theta+4\pi \sin\theta -4 \thh\sin 2\theta\right]
-D_0 \cos 2\theta\right\}
\nonumber \\
& &
+ {\rm higher}\ {\rm order}\ {\rm terms} \ ,
\label{eqn2h2}\\
& & \nonumber \\
F_3 & = &
-\frac{D_2\ln{x} \cos\theta}{2 x^2}
-\frac{1}{x^2}\left\{
\frac{D_2}{16\sin\theta}\left[4\thh \cos 2\theta +3 \sin 2\theta 
           +\pi \cos\theta (3 \sin^2\theta -2)\right]
 -D_0\cos\theta \right\}
\nonumber \\
& & 
+ {\rm higher}\ {\rm order}\ {\rm terms} \ ,
\label{eqn2f3}\\
& & \nonumber \\
F_4 & = &
\frac{D_1\sin\theta}{x} -\frac{D_1(\bar{f}_1+2\gamma\bar{\varphi}_1)
\sin\theta }{4x^2}
+ {\rm higher}\ {\rm order}\ {\rm terms} \ ,
\label{eqn2f4}
\end{eqnarray}
for $n=3$, 
\begin{eqnarray}
H_1 & = & 
\frac{D_2\sin 2\theta}{2 x^2} -\frac{D_3\sin 2\theta}{2 x^3}
+ {\rm higher}\ {\rm order}\ {\rm terms}   \ ,
\label{eqn3h1} \\
H_2 & = & 
1+ \frac{D_2 \cos 2\theta}{2x^2} +\frac{D_3 (9\sin^2\theta -2)}{6x^3}
+ {\rm higher}\ {\rm order}\ {\rm terms}   \ ,
\label{eqn3h2} \\
F_3 & = &
-\frac{D_2 \cos\theta}{2x^2}+\frac{\cos\theta}{18x^3}\left[
(9 D_1 D_2 -5 D_3)\sin^2\theta + 6 D_3 \right]
+ {\rm higher}\ {\rm order}\ {\rm terms}   \ ,
\label{eqn3f3} \\
F_4 & = & 
\frac{D_1 \sin\theta}{x}
-\frac{D_1 (\bar{f}_1+2\gamma\bar{\varphi}_1)\sin\theta}{4x^2} 
+\frac{\sin\theta}{12x^3}
\left[
\left(5 D_3 +3 D_1 D^{*} 
+15 D_4\right)\sin^2\theta -4 D_3 -12 D_4
\right]
\nonumber\\
& & 
+ {\rm higher}\ {\rm order}\ {\rm terms}   \ ,
\label{eqn3f4}
\end{eqnarray}
and for $n=4$, 
\begin{eqnarray}
H_1 & = & 
\frac{D_2 \sin 2\theta}{2 x^2}
+\frac{D_3 \ln{x} \sin 4 \theta}{4 x^4}
+\frac{D_3}{16x^4} 
\left[4\thh \cos 4\theta -\sin 4\theta -\pi\cos\theta(2-15\sin^2\theta)
\right]
+\frac{D_4\sin 4\theta}{8x^4}
\nonumber\\
& & 
+ {\rm higher}\ {\rm order}\ {\rm terms}   \ ,
\label{eqn4h1} \\
H_2 & = & 
1+\frac{D_2 \cos 2\theta}{2x^2}
+\frac{D_3 \ln{x}\cos 4\theta}{4 x^4}
+\frac{D_3}{16x^4}
\left[4\pi \sin\theta (2-5\sin^2\theta) -\cos 4\theta -4 \thh \sin 4\theta
\right]
+\frac{D_4\cos 4\theta}{8x^4}
\nonumber\\
& & 
+ {\rm higher}\ {\rm order}\ {\rm terms}   \ ,
\label{eqn4h2} \\
F_3 & = &
-\frac{D_2\cos\theta}{2x^2}
-\frac{D_1 D_2 \cos\theta \sin^2\theta}{2 x^3}
-\frac{D_3 \ln{x} \cos\theta \cos 2\theta}{4 x^4}
+ {\rm higher}\ {\rm order}\ {\rm terms}   \ ,
\label{eqn4f3}  \\
F_4 & = & 
\frac{D_1 \sin\theta}{x} - \frac{D_1 (\bar{f}_1+2\gamma\bar{\varphi}_1)
 \sin\theta}{4 x^2}
+\frac{\sin\theta}{4 x^3} 
\left[
\left(D_1 D^{*}-5 D_5 
\right)\sin^2\theta+4 D_5
\right]
+ {\rm higher}\ {\rm order}\ {\rm terms}   \ ,
\label{eqn4f4}
\end{eqnarray}
where $D_0$ -- $D_5$, $\bar{\varphi}_1$ and $\bar{f}_1$ are constants
and $D^{*} = [(\bar{f}_1+2\gamma\bar{\varphi}_1)^2 
+2(\bar{f}_1^2+\bar{l}_2-2 \bar{m}_2)]/4$ \cite{foot2}.
The gauge field functions $H_3$ and $H_4$ are related to 
the functions $F_3$ and $F_4$ by
\begin{equation}
H_3   = \sin\theta F_3 + \cos\theta F_4 \ , \ \ \
1-H_4 = \cos\theta F_3 - \sin\theta F_4  \ ,
\end{equation}
and the asymptotic solutions for an odd number of nodes are obtained 
from the asymptotic solutions for an even number of nodes by 
multiplying the functions $H_1$, $H_2$ and $H_4$ by minus one.
Note the explicit occurrence of $\theta$ in the expansions
for $n=2$ and 4.
For all $n$, the leading terms of the dilaton function are
\begin{equation}
\varphi = \frac{\bar{\varphi}_1}{x} 
+ \frac{\bar{\varphi}_3}{2x^3}(3 \cos^2\theta -1 )
+ {\rm higher}\ {\rm order}\ {\rm terms}   \ ,
\label{eqdil}
\end{equation}
and for the metric functions the equations (\ref{mfold})-(\ref{mlold})
hold \cite{foot1}.

We now discuss the properties of these asymptotic solutions,
assuming $x >> 1$.
First of all we note that along the $z$- and $\rho$-axis 
all functions fulfill the proper boundary conditions \cite{kk1,kk2,kk3,kk4}.
Furthermore, under the transformation $\theta \rightarrow \pi-\theta$
the gauge field functions $H_1$ and $F_3$ are odd 
while $H_2$ and $F_4$ are even, consequently
$H_3$ is odd while $H_4$ is even.

Concerning regularity of the asymptotic solutions along the symmetry axis,
we found, using the analysis of ref.~\cite{regu}, that local 
gauge transformations $\Gamma_{(z)}^{(n)}$ exist, which lead to locally regular
gauge potentials along the $z$-axis.
Thus, up to the order of the expansion the asymptotic solutions are regular.
 
For $n=2$ we checked that the contributions from the next orders 
($\ln{x}/x^3$ and $1/x^3$) do not change the regularity and symmetry properties 
of the asymptotic solutions. We conjecture that this is also true
for the corresponding higher order terms of the
$n=3$ and $n=4$ asymptotic solutions.

Since $\ln{x}/x^2$ and $\ln{x}/x^4$ can not be expanded as a power series in 
$1/x$ a naive construction of the asymptotic solutions as power series in 
$1/x$ does not lead to the proper solutions for $n=2$ and $n=4$, respectively.

Finally we note, 
that comparison of the asymptotic solution Eqs.~(\ref{eqn2h1})-(\ref{eqn2f4})
with the numerical solutions, e.~g. for $n=2$ and $k=1$, yields excellent 
agreement.


\begin{thebibliography}{000}
\bibitem{foot1} 
The asymptotic solution of \cite{kk3} for $n=2$ corresponds to
$D_2=0$, whereas the numerical solutions of
\cite{kk1,kk2,kk3,kk4,YMD} have $D_2 \ne 0$.
\end{thebibliography}
\end{document}